\documentclass[12pt]{article}
\usepackage[dvips]{graphicx}
\usepackage{hhline}
\title{Many-body corrections to the nuclear anapole moment II}
\author{Vladimir F. Dmitriev \thanks{E-mail: dmitriev@inp.nsk.su},
Vladimir B. Telitsin \thanks{E-mail: telitsin@inp.nsk.su} \\
\it Budker Institute ofNuclear Physics, 630090, Novosibirsk-90, Russia}

\begin{document}
\maketitle

\begin{abstract}
The contribution of many-body effects to the nuclear anapole
moment were studied earlier in \cite{DT97}. Here, more accurate
calculation of the many-body contributions is presented, which
goes beyond the constant density approximation for them used in \cite{DT97}.
The effects of pairing are now included. The accuracy of the short
range limit of the parity violating nuclear forces is discussed.
\end{abstract}
\vspace{2cm}
\begin{center}
BudkerINP 99 - 100
\end{center}
\newpage
\section{Introduction}
In the previous paper \cite{DT97},  referred here as I, the contribution of
the core nucleons to the anapole moment (AM), arising from the core
polarization effects, has been calculated in the random-phase approximation
(RPA). Recently, the first measurement of the AM of Cesium  has been
reported \cite{W97}. The immediate application of this measurement was the
attempt to deduce the pion-nucleon weak parity-non-conserving (PNC) coupling
constant $f_\pi$ \cite{H97}, \cite{FM97}.
The comparison of the measured anapole moment value with the one calculated
using a pure single-particle model leads to a value for $f_\pi$ that exceeds
by a factor 4 the value deduced from a parity violating measurement in
$^{18}F$ \cite{AH85}.  More sophisticated comparison was made in
Ref. \cite{WB98}, where the results from $^{18}F$, the AM of $^{133} Cs$,
and the upper bound for the AM of $^{205}Tl$ \cite{VM95} were used to deduce
both $f_\pi$ and $f_\rho$ PNC coupling  constants. The combination of the
coupling constants was found which satisfies both $^{18}F$ and $^{133}Cs$
experiments and which is barely in agreement with theory. These values,
however, are inconsistent with the constraint obtained from $^{205}Tl$
measurement. This situation, even independently on $^{18}F$ experiments,
raises the question how accurate is the theory of nuclear anapole moments.
Here we address this question and present more accurate calculation of the
many-body effects.

There are different contributions to the value of nuclear AM. The
most significant and the least model dependent contribution comes from the
valence nucleon. It is stable under variations of the single-particle
potential \cite{DKT94}. Within the single-particle model it is often
convenient to use the leading approximation (LA) \cite{FK80} producing the
simple analytical expression for the correction to the single-particle wave
function $\delta\psi$ due to the PNC potential. The LA was used in I
in calculations of the core nucleons contribution to the anapole moment.
This is another contribution which is not negligible. The accuracy of the LA
was estimated to be within 20\% in \cite{DKT94}. We found, that although this
is true for the upper level of the spin-orbit doublet, for the lower level
the difference between LA and the exact correction $\delta\psi$ can be
significant. This demands improving the calculation of the core nucleons
contribution.

Another phenomenon that has to be accounted for is the pairing. The reason
for this is that for $^{133}Cs$ nucleus there are single-particle states very
close to the Fermi surface. The transitions to these states produce  large
contribution to the core polarization. In the presence of pairing these
transitions will be strongly reduced, thus affecting the value of the
polarization effects and the nuclear anapole moment.

Finally, we discuss the constants of the short range effective PNC
interaction used in nuclear structure calculations. These constants were
obtained from the finite range meson exchange forces by comparing typical
matrix elements both finite range and short range interactions
\cite{DFST83, FKS84, FTS85}. The factors $W_\rho \approx 0.4$ and
$W_\pi \approx 0.16$ were found for $\pi$- and $\rho$-exchange forces. These
values, however, were obtained for $\alpha$-particle and the extension to
heavier nuclei should be checked separately. We found using the same
procedure that the factor $W_\pi$ is state dependent and it is different
for $^{133}Cs$ and $^{205}Tl$.

The paper is organized as follows. First, we discuss the accuracy of the LA.
Then, we present the exact calculations of the core polarization contribution
and demonstrate the necessity to include the pairing. Finally, we discuss
the constants of the short range PNC interaction.

\section{Core polarization contribution}

For completeness of the discussion let us to remind the basic equations
from I. There are several contributions to the AM arising from different
parts of the electromagnetic current. Apart from magnetization current which
gives the main contribution, there are contributions from the convection
current, the spin-orbit current and the contact current arising from
the momentum dependence of the corresponding parts of the Hamiltonian
\cite{DKT94}:
\begin{eqnarray} \label{e1} \nonumber
{\bf a}_s^a & = & \frac{\pi e \mu_a}{m}\mbox{\boldmath $r\times\sigma$}
\\ \nonumber
{\bf a}_{conv}^p & = & -\frac{\pi e}{m}\left\{{\bf p},
(r^2-\frac{Z}{A}<r^2_p>)\right\} \\
{\bf a}_{conv}^n & = & \frac{2\pi e}{m}\frac{Z}{A}<r^2_p>{\bf p}
\nonumber \\
{\bf a}_{ls}^p & = & -\pi
eU_{ls}^{pn}\rho_0\frac{N}{A}r^2\frac{df(r)}{dr}\mbox{\boldmath$ \sigma
\times  n$}
\nonumber \\
{\bf a}_{ls}^n & = & \pi
eU_{ls}^{np}\rho_0\frac{Z}{A}\frac{d(r^2f(r))}{dr}\mbox{\boldmath$\sigma
\times  n$}
\nonumber \\
{\bf a}^p_c & = & \frac{G_F}{\sqrt{2}}\frac{\pi e}{m}\rho_0
g_{pn}\frac{N}{A} r^2f(r)\mbox{\boldmath$\sigma$}
\nonumber \\
{\bf a}^n_c & = &
\frac{G_F}{\sqrt{2}}\frac{\pi e}{m}\rho_0 g_{np}\frac{Z}{A} r^2f(r)\mbox{
\boldmath$\sigma$}
\end{eqnarray}
Here $\{\;,\;\}$ is anticommutator, $<r^2_p>$ is the proton mean squared
radius, $\rho_0$ is the central nuclear density, $f(r)=\rho(r)/\rho_0$ -
nuclear density profile, and
$ U_{ls}^{pn} =U_{ls}^{np} = 134 \, MeV \cdot fm^5 $
is the proton-neutron constant of the effective spin-orbit
residual interaction \cite{DKT94}. $G_F$ is the weak interaction Fermi
coupling constant, $m$ is the proton mass, and $\mu_a$ is the nucleon magnetic
moment. The contact current contribution arises from the velocity dependence
of the effective nucleon-nucleon PNC forces \cite{DKT94,KHR}

The operators in Eq. (\ref{e1}) have 3 types of angular dependence. All of
them, except the convection ones, are spin dependent. All of them except
${\bf a}_c$ are of E1 type. The operators created by the contact current are
of M1 type. The convection terms in Eq. (\ref{e1}) differ from the one
used in I and \cite{DKT94}. Here we accounted for the recoil correction
by subtracting the part related to the motion of a nucleus as a whole.

As in I, we use for the radial dependence of the effective operators related
to Eq. (\ref{e1}) the notation $V[{\bf a}_i]({\bf r})$.  The RPA equations for
the effective radial fields $V[{\bf a}_i]({\bf r})$ are
(see Eqs. (17),(18) in I)
\begin{equation} \label{e2}
V = V_0([{\bf a}_i]) + F_s\,A\,V,
\end{equation}
\begin{equation} \label{e3}
\delta V = F_w\,AV + F_s\,\delta AV + F_s\,A\delta V.
\end{equation}
Here $F_w$ is the PNC nucleon-nucleon interaction
\[
F_w  =\frac{G}{\sqrt 2} \frac{1}{4m}
\sum_{a,b}\left(\{(g_{ab}\mbox{\boldmath$\sigma$}_a -
g_{ba}\mbox{\boldmath$\sigma$}_b)\cdot({\bf p}_a - {\bf p}_b),
\delta({\bf r}_a -
{\bf r}_b)\}\right.
\]
\begin{equation}\label{e4}
\left. +\ g^{\prime}_{ab}\mbox{\boldmath$[\sigma_a\times\sigma_b]$}
\cdot\mbox{\boldmath$\nabla$}\delta({\bf r}_a - {\bf r}_b)\right),
\end{equation}
and $F_s$ is the residual spin-spin interaction
\begin{equation} \label{e5}
F_s(ab) = C\,\left( g_0 +
g'_0 \mbox{\boldmath$\tau_a \cdot\tau_b$} \right)\mbox{\boldmath
$\sigma_a \cdot\sigma_b$} \delta ({\bf r}_a - {\bf r}_b).
\end{equation}
As in I, we use $C=300\;MeV\cdot fm^3$, $g_0 = 0.6$, and $g_0'= 1$.
Averaging the interaction Eq.(\ref{e4}) over core particles we obtain
a single-particle PNC potential
\begin{equation}\label{e4'}
W_a({\bf r}) = \frac{G}{\sqrt{2}}\frac{g_a\rho_0}{2m}\{(\mbox{\boldmath
$\sigma$} \cdot{\bf p}),f(r)\},
\end{equation}
where $\rho_0$ is the central nuclear density, $f(r)$ is the nuclear density
profile and the constants $g_a$ are related to the interaction constants
$g_{ab}$ (see Eq. (4) in I). In the absence of pairing, the static
particle-hole propagator $A({\bf r},{\bf r}')$ has the form
\begin{equation}\label{e6}
A({\bf r},{\bf r}') = \sum_{\nu \nu'}\psi_\nu({\bf r}) \psi^\dagger_\nu
({\bf r}')\frac{n_\nu -n_{\nu'}}{\epsilon_\nu -\epsilon_{\nu'}}\psi_{\nu'}
({\bf r}')\psi^\dagger_{\nu'}({\bf r}),
\end{equation}
where $n_\nu$ are the occupation numbers, $\epsilon_\nu$ and $\psi_\nu({\bf
r})$ are the energies and the wave functions of the single-particle levels.
The Feynman diagrams corresponding to Eq.(\ref{e2},\ref{e3}) are shown in
Fig.1.
\begin{figure} [h]
\includegraphics[width=12 cm]{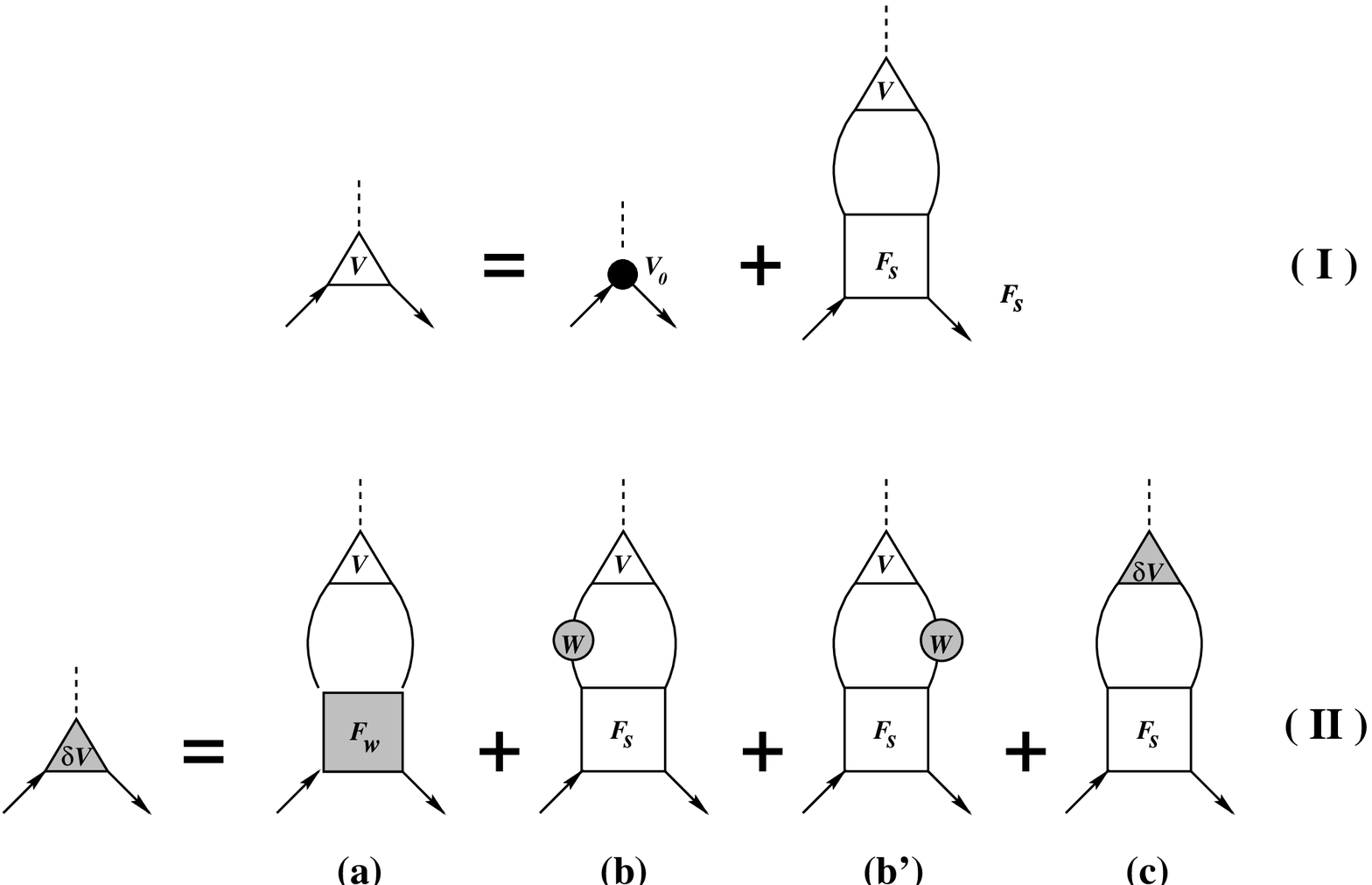}
\caption{Feynman diagrams corresponding to Eqs.(2),(3).
\protect \\
I. The anapole moment renormalization diagram describing the core polarization
effects. The filled circle is the bare anapole operator. The open triangle is
the dressed one. The open square is the spin-spin residual strong interaction.
\protect
\\
II. Additional contribution to the anapole moment due to parity violation in
the core states.
a) - the direct contribution of the two-particle effective PNC interaction.
b) and b') - the effective contribution due to parity violation in the
core states.
c) - the renormalization diagram. The shaded circle is the single-particle
PNC interaction. The shaded square is the two-particle effective PNC
interaction.}
\end{figure}
Eq.(\ref{e2}), which is shown in the top diagrams (I) in Fig.1, describes the
usual RPA renormalization of the bare operator $V_0[{\bf a}_i]({\bf r})$.
The next Eq.(\ref{e3}), which is shown in the bottom diagrams (II) in Fig.1,
describes  an additional contribution from the core nucleons arising both
from the direct P-odd nucleon-nucleon interaction $F_w$, see Fig.1 (II a),
and from the P-even residual interaction via admixture of opposite parity
states to the wave functions of the core nucleons,see Fig.1 (II b), (II b').
The last term in Eq.(\ref{e3}), see Fig.1 (II c), is responsible for the
renormalization of these contributions.

It is worth mentioning that the correction $\delta V$ has the parity opposite
to $V$. The change in parity happens because $\delta V$ is created from $V$
by the interaction that does not conserve parity.
While AM is E1 type operator, the correction $\delta V$ is M1 type.
For this reason the renormalization due to the core polarization will be
different for $V$ and $\delta V$, since different transitions are involved in
the kernels of integral equations (\ref{e2}) and (\ref{e3}). Eq.(\ref{e2})
describes also the renormalization of the operators ${\bf a}_c$. They don't
have $\delta V$ since they themselves are of first order in the weak PNC
interaction. The AM value is given by the sum of all these terms:
\begin{equation} \label{e7}
a_i = \langle \delta \psi|V[{\bf a}_i]|\psi\rangle
+ \langle\psi|V[{\bf a}_i]|\delta \psi\rangle
+ \langle\psi|\delta V[{\bf a}_i]|\psi\rangle,
\end{equation}
where the index i means here $s$, $conv$ or $ls$. For the contact contribution
we have
\[
a_c = \langle \psi|V[{\bf a}_c]|\psi \rangle.
\]
\subsection{Leading approximation accuracy}
Eq.(\ref{e3}) was solved in I using LA in calculations of $\delta A$.  The
accuracy of LA was estimated in \cite{DKT94} within 20\%. This estimate was
maid for odd valence nucleons in a set of nuclei. Occasionally, all the
valence nucleon levels under discussion were the upper levels of the
spin-orbit doublet. The typical difference between exact $\delta R(r)$ and
$rR(r)$ in LA for the upper level of the spin-orbit doublet is shown in the
left plot of Fig.2.
\begin{figure}[h]
\includegraphics[width=6cm]{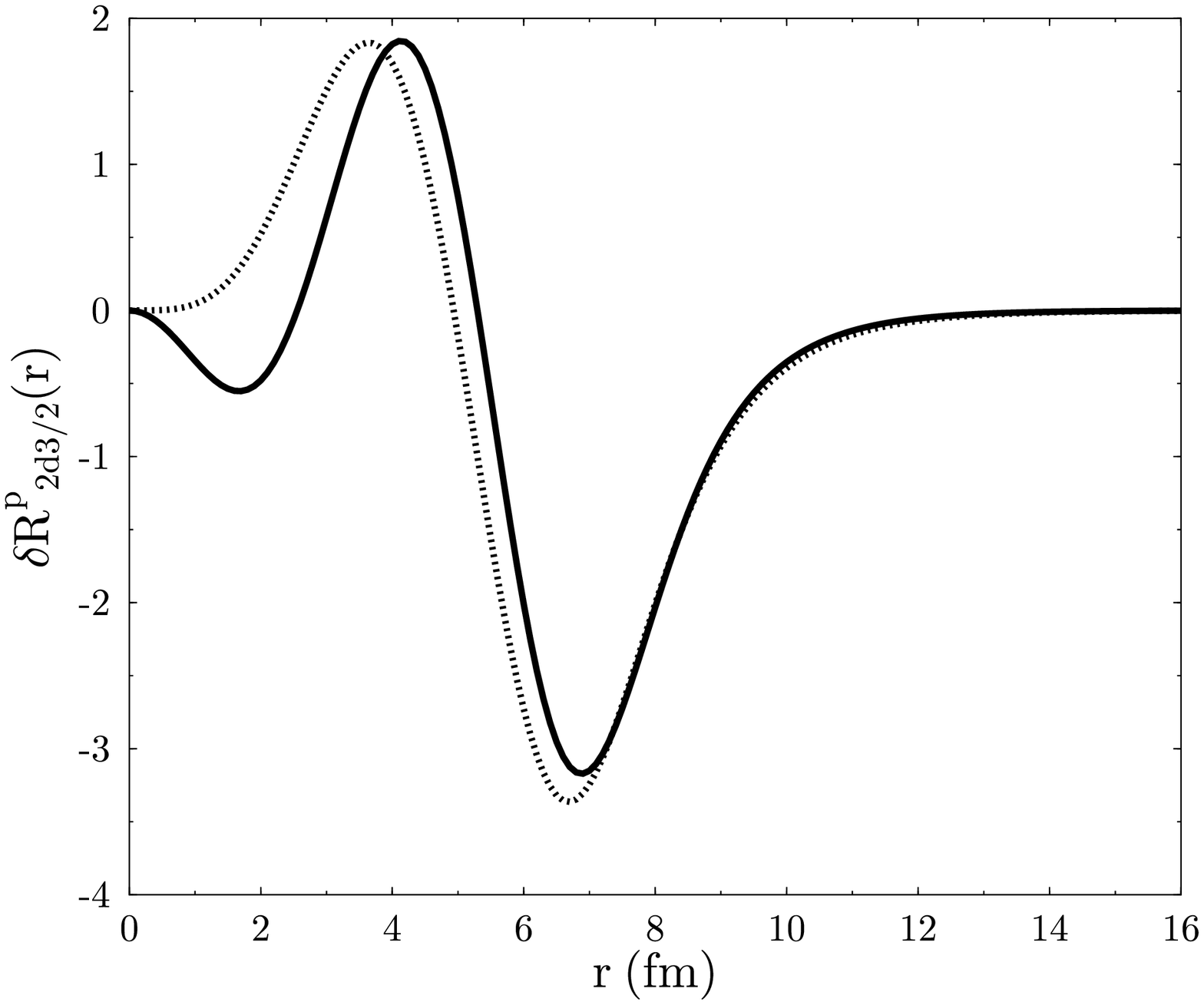}
\includegraphics[width=6cm]{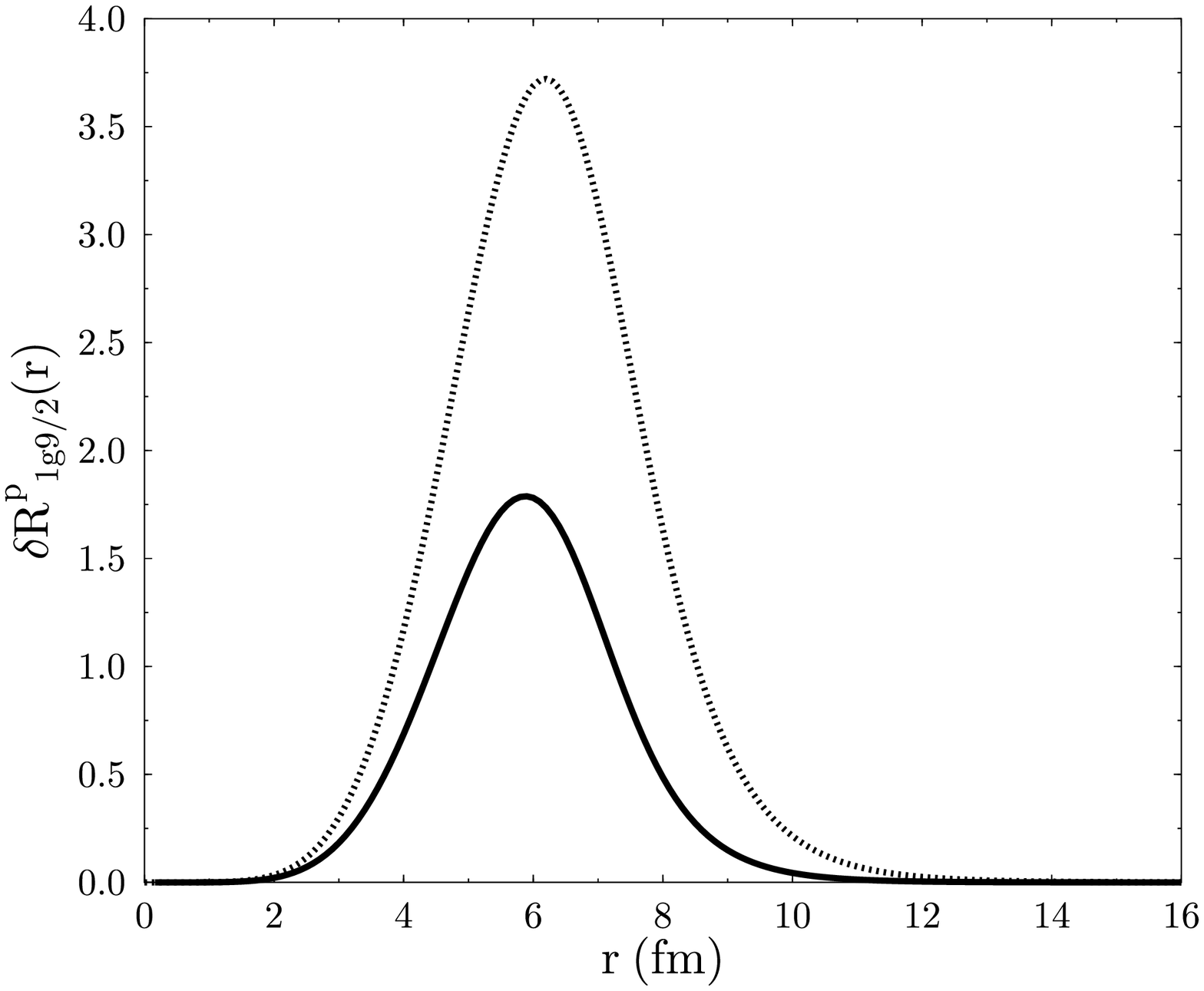}
\caption{Admixture of the opposite parity $\delta R(r)$ for different
components of the spin-orbit doublets. Left panel - 2d$_{3/2}$ level, right
panel - 1g$_{9/2}$ level. LA is shown by the dashed lines. Full lines show
the exact solution.}
\end{figure}
However, the sum in $\delta A$ goes over all  states. For the lower states of
the doublet the difference in the peak heights can reach a factor 2 as one
can see in the right plot in Fig.2. In all cases the exact correction and
the LA are peaked near the surface. However, the ratio of the peak heights
differs considerably for the upper and the lower states of spin-orbit
doublets.

The systematic study of this ratio is presented in Fig. 3.
\begin{figure}[h]
\includegraphics[height=8cm,width=12cm]{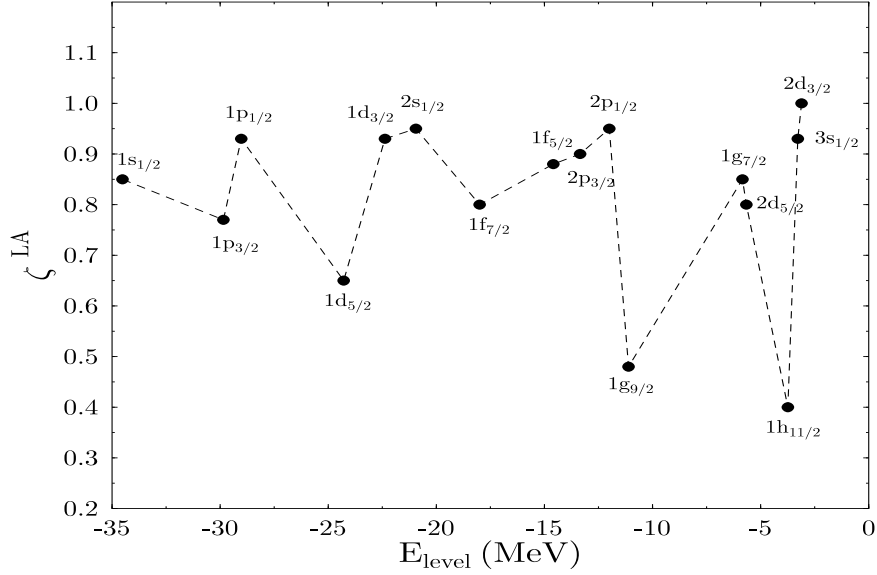}
\caption{Ratio $\zeta$ of the maximum in exact $\delta R(r)$ to the maximum in
 $\delta R(r)$ found in LA for a set of the proton single-particle orbitals.}
\end{figure}
 Here we plotted the ratio of the peak
heights of the exact $\delta R(r)$ and $rR(r)$ in the LA for a set of proton
states in the Woods-Saxon potential.  While for the states with $j=l-1/2$ the
ratio remains close to 1, for the states with $j=l+1/2$ the ratio goes down
with increasing orbital angular momentum $l$.  For the state $1h_{11/2}$ the
exact correction $\delta R(r)$  differs more than by factor 2 from the
correction $rR(r)$ calculated using LA.  From these results we see that
the estimated accuracy 20\% in calculation of $\delta A$ cited in I is too
optimistic and more accurate calculation is necessary.

\subsection{Exact $\delta A$}

Let us start with the expression $F_sAV$ in Eq.(\ref{e2}). In coordinate
space it can be presented as follows
\[
(F_sAV[{\bf a}_i])_b({\bf r}) = Cg^{bc}_0\mbox{\boldmath $\sigma$}
{\bf X}_c({\bf r}),
\]
where $C$ and $g_{bc}$ are the parameters of the residual interaction
(\ref{e5}), and
\begin{equation}\label{e8}
{\bf X}({\bf r}) = \sum_{\nu\nu'}\frac{n_\nu - n_{\nu'}}{\epsilon_\nu -
\epsilon_{\nu'} +
\omega} \int\;d^3r'\;\psi^\dagger_\nu ({\bf r})\mbox{\boldmath
$\sigma$}\psi_{\nu'}({\bf r})\psi^\dagger_{\nu'}({\bf r}')V({\bf r}')
\psi_\nu ({\bf r}').
\end{equation}
 In the following derivation we keep the external
frequency $\omega$. It will be put to zero in the final result. Introducing
the single-particle Green's function
\[
G({\bf r\;r}'|E)= \sum_\nu \frac{\psi_\nu({\bf r})\psi^\dagger_\nu
({\bf r}')}{E-\epsilon_\nu}
\] \\
we can rewrite the Eq.(\ref{e8}) in the following form
\[
 {\bf X}({\bf r}) =
\sum_\nu n_\nu \, \int\;d^3r\;\left\{\psi^\dagger_\nu({\bf r})
\mbox{\boldmath $\sigma$}
G({\bf r\;r}'|\epsilon_\nu +\omega)V({\bf r}')\psi_\nu({\bf r}')\right.
\]
\begin{equation} \label{e9}
+\left.\psi^\dagger_\nu({\bf r}')V({\bf r}')G({\bf r}'\;
{\bf r}|\epsilon_\nu-\omega)\mbox{\boldmath $\sigma$}
\psi_\nu({\bf r})\right\}.
\end{equation}

The correction to Eq.(\ref{e9}) arising from the single-particle PNC potential
can be obtained using the correction to the single-particle wave
function $\psi_\nu ({\bf r})$ defined in I:
\begin{equation}\label{e10}
\delta\psi_\nu ({\bf r}) = -\imath\xi_a(\mbox{\boldmath $\sigma$}{\bf n})
\delta R_\nu (r)\Omega_\nu ({\bf n})
= \imath\xi_a \delta R_\nu (r)\Omega_{\tilde{\nu}}({\bf n}),
\end{equation}
where
\[
\xi_a= \frac{G}{\sqrt{2}}g_a\rho_0.
\]
We used here the relation
\[
\Omega_{j\tilde{l}m}({\bf n}) = - (\mbox{\boldmath $\sigma$}{\bf
n})\Omega_{jlm}({\bf n}),
\]
where $\tilde{l} = 2j-l$. For the correction
$\delta{\bf X}({\bf r})$ we find then
\[
\delta{\bf X}({\bf r}) = \delta{\bf
X}_1({\bf r}) +\delta{\bf X}_2({\bf r}),
\]
where
\[
\delta{\bf X}_1({\bf r}) = \imath\xi\sum_{\nu\nu'}n_\nu\int r'^2 dr'
\]
\[
\times\left\{(\Omega^ \dagger_\nu ({\bf n})\mbox{\boldmath
$\sigma$}\Omega_{\nu'}({\bf n}))G_{\nu'}(r,r'|\epsilon_\nu+ \omega)
<\nu'|V({\bf r}')|\tilde{\nu}>\delta R_\nu(r')R_\nu(r)\right.
\]
\[
-\ (\Omega^\dagger_{\tilde{\nu}}({\bf n})\mbox{\boldmath $\sigma$}
\Omega_{\nu'}({\bf n}))G_{\nu'}(r,r'|\epsilon_\nu
+ \omega)<\nu'|V({\bf r}')|\nu>\delta R_\nu(r) R_\nu (r')
\]
\[
+\ \delta R_\nu (r)R_\nu (r')<\nu|V({\bf r}')|\nu'>(\Omega^\dagger_{\nu'}
({\bf n})\mbox{\boldmath $\sigma$}\Omega_{\tilde{\nu}}({\bf n}))
G_{\nu'}(r'r|\epsilon_\nu -\omega)
\]
\begin{equation}\label{e11}
\left.\ \ \ \ \ \ \ \ -\ R_\nu (r)\delta R_\nu (r')
<\tilde{\nu}|V({\bf r}')|\nu'>
(\Omega^\dagger_{\nu'}({\bf n})\mbox{\boldmath $\sigma$}\Omega_\nu({\bf n}))
G_{\nu'}(r'r|\epsilon_\nu-\omega)\right\},
\end{equation}
and
\[
\delta{\bf X}_2({\bf r}) = \sum_\nu n_\nu\int d^3r'
\left\{ \psi^\dagger_\nu ({\bf r})
\mbox{\boldmath $\sigma$}\delta G({\bf r\,r}'|\epsilon_\nu +\omega)V({\bf r}')
\psi_\nu({\bf r}')\right.
\]
\begin{equation}\label{e12}
\left.+\ \psi^\dagger_\nu ({\bf r}')V({\bf r}')
\delta G({\bf r\,r}'|\epsilon_\nu - \omega)
\mbox{\boldmath $\sigma$}\psi_\nu ({\bf r})\right\}.
\end{equation}
The angular brackets in Eq.(\ref{e11}) denote the matrix elements in angular
and spin variables only. In the following calculations for spherical nuclei,
it is convenient to separate the angular variables by  expanding both
$\delta{\bf X}_1({\bf r})$ and $\delta{\bf X}_2({\bf r})$ in a set of the
vector spherical harmonics ${\bf Y}^L_{JM}({\bf n})$,
\begin{equation}\label{e13}
\delta{\bf X}_i({\bf r}) = \sum_{L} X_{i}^L(r){\bf Y}^L_{JM}({\bf n}).
\end{equation}
In Eq.(\ref{e13}) the total angular momentum $J$ is fixed, $J=1$, and the sum
goes over $L$ only. Multiplying Eqs. (\ref{e11} -\ref{e12}) by
$ {\bf Y}^{L\dagger}_ {JM}({\bf n})$, integrating over angles and separating
the dependence on all angular momentum projections $M$ via Wigner-Eckart
theorem, we obtain for the reduced functions of $r$
\[
X^L_1(r)= \frac{\imath\xi}{3}\sum_{\nu\nu'}n_\nu\int
r'^2\, dr' \left\{ \delta R_\nu (r) R_\nu (r')(\nu ||V(r')||\nu')
(\tilde{\nu}||T^L_1||\nu')
\right.
\]
\[
\left. -\ \delta R_\nu (r')R_\nu (r)(\tilde{\nu}||V(r')||\nu')
(\nu ||T^L_1||\nu') \right\}
\]
\begin{equation}\label{e14}
\times \ \bigl(G_{\nu'}(r,r'|\epsilon_\nu +\omega)
+ G_{\nu'}(r,r'|\epsilon_\nu-\omega) \bigr).
\end{equation}
Here, double vertical lines mean the reduced angular matrix elements. The
tensor operators $T^L_{JM}$ are defined as in I:
\begin{equation}\label{e14'}
T^L_{JM} = \{\sigma \otimes Y_L\}_{JM}
\end{equation}

Similar expression can be obtained for $X^L_2(r)$:
\[
X^L_2(r)= \frac{1}{3}\sum_{\nu\nu'} n_\nu (\nu'||T^L_1||\nu ) R_\nu (r)
\]
\begin{equation}\label{e15}
\times \int r'^2dr'\bigl(\delta G_{\nu'\tilde{\nu'}}
(r,r'|\epsilon_\nu + \omega)
+\delta G_{\nu'\tilde{\nu'}}(r,r'|\epsilon_\nu - \omega) \bigr)
(\tilde{\nu'}||V(r')||\nu )R_\nu (r').
\end{equation}
In Eqs.(\ref{e12},\ref{e15}) $\delta G$ is the correction to the Green's
function of the Schr\"odinger equation for single-particle orbitals due to
the PNC potential Eq.(\ref{e4'}). In first order the correction is
\[
\delta G({\bf r},{\bf r}') = \int\,d^3x G({\bf r},{\bf x})W({\bf x})
G({\bf x},{\bf r}')
\]
Separating the angular dependence we obtain for the radial correction
\begin{equation}\label{e16}
\delta G_{\nu\tilde{\nu}}(r,r')=\frac{\imath\xi}{2m}\int_0^\infty x^2dx\, f(x)
\left(G_\nu (r,x)(\frac{\begin{array}{c}{\leftarrow \rightarrow}\\ \partial
\end{array}}{\partial x}-\frac{2k}{x})
G_{\tilde{\nu}}(x,r') \right).
\end{equation}
Here $k=(l-j)(2j+1)$. For obvious reason the, PNC potential couples only the
states with the angular momenta $l$ and $\tilde{l}$.

The final equations for the radial functions $v[{\bf a}_i](r)$ and
$\delta v_L[{\bf a}_i](r)$ are as follows
\begin{equation}\label{e18}
v^a[{\bf a}_i](r) - \sum_{b=p,n}g_0^{ab}\int_0^\infty r'^2dr'
\,A^b_{11}(r,r')v^b[{\bf a}_i](r') = v^a_0[{\bf a}_i](r);
\end{equation}
\begin{equation}\label{e19}
\delta v_L^a[{\bf a}_i](r) - \sum_{b=p,n}g_0^{ab}\sum_{L'=0,2}\int_0^\infty
r'^2dr'\,A_{LL'}^b(r,r')\delta v_{L'}^a[{\bf a}_i](r') =
\delta v_L^{0a}[{\bf a}_i](r),
\end{equation}
where the radial particle-hole propagator $A_{LL'}^b(r,r')$ is
\[
A_{LL'}^b(r,r') = \frac{1}{3}C\sum_{jlnj'l'}n^b_{jln}\langle jl||T^L_1||j'l'
\rangle\langle jl||T^{L'}_1||j'l'\rangle^* R_{jln}(r)R_{jln}(r')
\]
\begin{equation}\label{e20}
\times\ \bigl(
G_{j'l'}(r,r';\epsilon_{jln}+\omega)+G_{j'l'}(r,r';\epsilon_{jln}-\omega)
\bigr),
\end{equation}
with $\omega \rightarrow 0$. Here $C$ is the normalization constant of the
residual interaction Eq.(\ref{e5}) and $G_{jl}(r,r';\epsilon)$ is Green's
function of the radial Schr\"odinger equation.

Eqs.(\ref{e18}),(\ref{e19}) are similar for all anapole operators
Eq.(\ref{e1}),
except for the convection current operator which will be treated separately.
The right-hand-side of Eq.(\ref{e18}) for the spin current anapole operator is
\begin{equation}\label{e21}
v^a_0[{\bf a}_s](r) =-\imath\sqrt{\frac{8\pi}{3}}\frac{\pi e\mu_a}{m}r.
\end{equation}
The transition from the Cartesian vector product in Eq.(\ref{e1}) to the
tensor operator $T^1_{1M}$ results in the factor $-\imath\sqrt{8\pi/3}$.
The factor is common for the spin and spin-orbit anapole operators.
The radial dependence follows Eq.(\ref{e1}).
The right-hand-side of Eq.(\ref{e19}) is given by
\begin{equation}\label{e24}
\delta v^{0a}_L[{\bf a}_i](r) =
C\sum_{b=p,n}g_0^{ab}(X_{1,b}^L(r)+X_{2,b}^L(r)).
\end{equation}

In Eq.(\ref{e24}) we omit the direct contribution of the two-particle
effective PNC interaction corresponding to the diagram (IIa) in Fig.1.
According to the estimates made in Ref.\cite{FKS85}
this contribution is small, it does not have the factor $A^{2/3}$.
The estimate in \cite{FKS85} was obtained with harmonic oscillator wave
functions. We calculated the diagram (IIa) for $V^0[{\bf a}_s]$ from
Eq.(\ref{e1}) with the Woods-Saxon wave functions. For $^{133}$Cs nucleus we
obtained the contribution
\[
 \delta\kappa_{F_w} =
(-4.1g_{pp} +0.1g_{pn} +4.9g_{np} -1.6g'_{pn})\cdot 10^{-3}
 = -0.015
\]
for the "best values" of the coupling constants \cite{DDH}. This number should
be compared with the single-particle value of $\kappa_s = 0.32$, and it gives
the contribution less than $5\%$. Similar calculations give
$\delta\kappa_{F_w} =0.003$ for $^{205}Tl$,  $\delta\kappa_{F_w} = 0.023$
for $^{209}Bi$, and $\delta\kappa_{F_w} = 0.0085$ for $^{207}Pb$.
All these values are in agreement with the estimates of Ref.\cite{FKS85}.

The convection current anapole operator Eq.(\ref{e1}) expressed via tensor
operators has the form
\[
(a^p_{conv})_M = -\imath\sqrt{\frac{8\pi}{3}}\frac{\pi e}{m}
\left(\frac{r^2-\frac{Z}{A}<r_p^2>}{2r}\{ \right.
{\bf l},{\bf Y}^1_{1M}\}
\]
\begin{equation}\label{e22}
-\ \frac{1}{\sqrt{8}}\{\frac{\partial}{\partial r}
\left. +\frac{1}{r},r^2 -\frac{Z}{A}<r_p^2>\}Y_{1M}\right).
\end{equation}
\[
(a^n_{conv})_M = \imath\sqrt{\frac{8\pi}{3}}\frac{\pi e}{m}\frac{Z}{A}<r_p^2>
\left(\frac{1}{2r}\{
{\bf l},{\bf Y}^1_{1M}\} -
\frac{1}{\sqrt{2}}(\frac{\partial}{\partial r}
+\frac{1}{r})Y_{1M}\right).
\]
The operators Eq.(\ref{e22}) are non-local and do not include spin. For
the last reason they are not renormalized directly by the residual interaction
Eq.(\ref{e5}). However, due to the spin-orbit potential a new spin dependent
contribution to AM is generated in first order in the residual interaction
Eq.(\ref{e5}). Its tensor structure is given by the same tensor operator as
that for
the spin current AM ${\bf a}_s$. The radial dependence of this polarization
contribution is given by
\[
v^{a0}_{pol}(r) = \frac{C}{3}\sum_{b=p,n}g_0^{ab}\sum_{\nu\nu'}n_{\nu}R_\nu (r)
(\nu ||T^1_{1}||\nu')^*
\]
\begin{equation}\label{e23}
\times \int_0^\infty r'^2dr'R_\nu (r')(\nu ||a^b_{conv}||\nu')
\bigl(G_{\nu'}(r',r|\epsilon_\nu +\omega)+G_{\nu'}(r,r'|\epsilon_\nu-\omega)
\bigr),
\end{equation}
where $\omega \rightarrow 0$. This operator undergoes the usual renormalization
described by Eq.(\ref{e18}) producing in the next orders in the residual
interaction the renormalized radial operator $v_{pol}^a(r)$. Although the
direct contribution of the convection current to AM is small, the polarization
contribution Eq.(\ref{e23}) is larger and has to be included. Total operator
generated by the convection current has the following structure
\begin{equation} \label{e23'}
v^b[{\bf a}_{conv}](r) = a^b_{conv} + v^b_{pol}(r)T_1^1,
\end{equation}
where $a^b_{conv}$ is given by Eq.(\ref{e22}) and $v^b_{pol}(r)$ is the dressed
polarization operator Eq.(\ref{e23}). The tensor operator $T_1^1$ is defined by
Eq.(\ref{e14'}) and we omit the total momentum projection.
This very operator Eq.(\ref{e23'}) should
be used in the Eq.(\ref{e24}) for the right-hand-side of Eq.(\ref{e19}).

The values of the AM resulting from the solutions of Eqs.(\ref{e18}),
(\ref{e19}) are summarized in Table 1 and Table 2 for $^{133}Cs$ and $^{205}Tl$
nuclei. For comparison, in the row labeled by I we listed the results of
previous calculation with the use of LA. The results of present calculations
are listed in the row labeled by II.

The value cited in column $V -V_0$ is the renormalization of the
single-particle (column s.p.) value due to the core polarization by
Eq.(\ref{e3}). One can see that the polarization contribution is about half
of the single-particle one and it has an opposite sign compared to the
single-particle contribution for all
spin dependent operators. This is in accordance with the repulsive nature
of the spin-spin residual interaction Eq.(\ref{e5}). In this case the core
produces a screening of the valence nucleon spin.

The contribution of the convection current in the Tables 1 and 2 was listed
together with the polarization effects discussed above. The single-particle
value is small compared to the magnetization current contribution. Let us note
that it differs from the corresponding value cited in I. The difference,
although small, comes from the center of-mass-motion that was not excluded
in I. It is interesting to note that for the convection
current, the polarization contribution has the same sign as the
single-particle one. This is in contrast to the spin case where the
polarization contribution has always the opposite sign compared to the
single-particle value. The difference between these two cases lies in the
polarization loop which is non-diagonal for the convection current case.
It has the spin operator in one vertex and the convection part of the AM
(see Eq.(\ref{e1})) in the other. For such non-diagonal loop the sign is not
fixed.

The main conclusion that can be drawn from the present results is that the LA
overestimates the contribution of $\delta V$ approximately by a factor 2.
The summed value $\kappa_{tot}$ has decreased by 50\%, as compared to its
single-particle value.  Thus, the effect of the core polarization is
considerable.

\section{Pairing effects}

In general, the pairing effects are important only for the transitions near
the Fermi surface. Therefore, they can be neglected in calculations of
$\delta\psi$ and in the polarization loop in Eq.(\ref{e18}) where the
transitions to at least next shell are involved. The situation is different
for Eq.(\ref{e19}) where the transitions within the open shell are allowed.
Good example is $^{133}Cs$ where the
polarization loop includes the transitions between the states with the
excitation energy about one hundred KeV. Transitions to such close levels
contribute significantly into polarization loop creating large response in
$\delta V$.  These
transitions will apparently be suppressed by pairing correlations. For these
reason the polarization loop in Eq.(\ref{e19}) should be modified in order to
include pairing.

The modification is straightforward and can be done similarly to \cite{PS88}.
The particle-hole propagator for a T-odd channel in the presence of pairing
has the form
\begin{equation}\label{e25}
A({\bf r},{\bf r'})=-\sum_{\nu\nu'}\frac{(u_\nu v_{\nu'}-v_\nu u_{\nu'})^2
\psi_\nu({\bf r})\psi^\dagger_{\nu'}({\bf r})\psi_{\nu'}({\bf r'})\psi^\dagger
_\nu({\bf r'})}{E_\nu + E_{\nu'}},
\end{equation}
where $u_\nu$ and $v_\nu$ are the Bogolyubov factors, and
$E_\nu=\sqrt{(\epsilon_\nu - \mu)^2 + \Delta^2}$ where
$\Delta$ and $\mu$ are the pairing gap and
the chemical potential. Below, we shall neglect slight state dependence of
the pairing gap and put $\Delta = const$.
Let us divide the single-particle space into 3 regions.
Let $P_\nu$ be a projection operator onto the region near the Fermi
surface, where pairing produces a significant effect.
It is convenient to include here all bound states above the Fermi surface.
Let $Q_\nu$ be a projection operator onto the hole states below the pairing
region. In this region we neglect the pairing effects. The states in
continuum will be projected by $1-P_\nu-Q_\nu$. Introducing the identity
\[
1 = (P_\nu + Q_\nu +1 -P_\nu - Q_\nu)(P_{\nu'} + Q_{\nu'} +1 -P_{\nu'} -
Q_{\nu'})
\]
into Eq.(\ref{e25}) we obtain the expression for the particle-hole loop as a
sum of 7 following terms
\[
A({\bf r},{\bf r'})=-\sum_{\nu\nu'}\psi_\nu({\bf r})\psi^\dagger_{\nu'}({\bf
r})\psi_{\nu'}({\bf r'})\psi^\dagger_\nu({\bf r'})\times \left\{
 \frac{(u_\nu v_{\nu'}-v_\nu u_{\nu'})^2 }{E_\nu + E_{\nu'}}P_\nu P_{\nu'}
 \right.
\]
\[
+\frac{u_\nu^2 }{E_\nu -\epsilon_{\nu'}+\mu}P_\nu Q_{\nu'}+ \frac{u_{\nu'}^2
}{E_{\nu'} -\epsilon_\nu+\mu}P_{\nu'} Q_\nu
\]
\[
+\frac{v_\nu^2 }{E_\nu +\epsilon_{\nu'}-\mu}P_\nu(1-P_{\nu'}- Q_{\nu'})+
\frac{v_{\nu'}^2}{E_{\nu'} +\epsilon_\nu-\mu}P_{\nu'}(1-P_\nu- Q_\nu)
\]
\begin{equation}\label{e26}
\left. -\frac{1}{\epsilon_{\nu'} -\epsilon_\nu}Q_{\nu'}(1-P_\nu- Q_\nu)-
\frac{1}{\epsilon_\nu -\epsilon_{\nu'}}Q_\nu(1-P_{\nu'}- Q_{\nu'})\right\}.
\end{equation}
In Eq.(\ref{e26}) we put $\Delta =0$ in all terms containing $Q_\nu$. The sum
over the whole single-particle space is present in the last 4 terms containing
the unit projection operator. Combining these terms together we can present
Eq.(\ref{e26}) as a sum of three different terms
\[
A({\bf r},{\bf r}') = A_Q({\bf r},{\bf r}') + A_P({\bf r},{\bf r}') +
A_\theta({\bf r},{\bf r}'),
\]
where
\begin{equation}\label{e26'}
A_Q({\bf r},{\bf r}') =\sum_\nu Q_\nu\{\psi_\nu({\bf r})\psi_\nu^\dagger({\bf
r}' )G({\bf r}',{\bf r}|\epsilon_\nu)+G({\bf r},{\bf r}'|\epsilon_\nu)\psi_\nu
({\bf r}')\psi_\nu^\dagger({\bf r})\}
\end{equation}
includes the transitions from the deep holes,
\begin{equation}\label{e26''}
A_P({\bf r},{\bf r}')=-\sum_\nu P_\nu v_\nu^2\{\psi_\nu({\bf r})
\psi_\nu^\dagger({\bf r}' )G({\bf r}',{\bf r}|\mu - E_\nu)
+ G({\bf r},{\bf r}'|\mu - E_\nu)\psi_\nu({\bf r}')\psi_\nu^\dagger({\bf r})\}
\end{equation}
includes the transitions from the pairing region, and
\begin{equation}\label{e26'''}
A_\theta({\bf r},{\bf r}') =-\sum_{\nu\nu'}\psi_\nu({\bf r})
\psi_\nu^\dagger({\bf r}' )\Theta_{\nu\nu'}\psi_\nu({\bf r}')
\psi_\nu^\dagger({\bf r})
\end{equation}
with
\[
\Theta_{\nu\nu'}=[\frac{(u_\nu v_{\nu'}-v_\nu u_{\nu'})^2}{E_\nu + E_\nu'}
-\frac{v_\nu^2}{E_\nu -\epsilon_{\nu'} +
\mu}-\frac{v_{\nu'}^2}{E_{\nu'}-\epsilon_\nu + \mu}]P_\nu P_{\nu'}
\]
\[
+ [
\frac{u_\nu^2}{E_\nu - \epsilon_{\nu'} + \mu} - \frac{v_\nu^2}{E_\nu
+\epsilon_{\nu'} - \mu} + \frac{1}{\epsilon_{\nu'}-\epsilon_\nu}]P_\nu Q_{\nu'}
\]
\[
+ [\frac{u_{\nu'}^2}{E_{\nu'}-\epsilon_\nu + \mu} -\frac{v_{\nu'}^2}{E_{\nu'} +
\epsilon_\nu -\mu} +\frac{1}{\epsilon_\nu - \epsilon_{\nu'}}]P_{\nu'} Q_\nu.
\]
includes the transitions from the pairing regions to the region of deep holes.
 In all these terms the sum goes over a
finite range of the single-particle space and can be calculated directly.

Strictly speaking, in the presence of pairing the particle-particle channel
should be added to Eq.(\ref{e23}) and the equation for $V$ has to be modified.
With the particle-particle channel a new interaction $F^\xi$ appears and
the total number of equations doubles. The parameters of the interaction
$F^\xi$ can be found from masses of the nuclei that differ by two protons or
two neutrons \cite{ST65}. This procedure, however, determines the
spin-independent part of the interaction $F^\xi$. The spin-independent
interaction in the short-range approximation does not renormalize  the
spin-dependent operators. As for the spin-dependent part of $F^\xi$, its
magnitude remains unknown up to now and for this reason we did not include
it in our equations.

For the right-hand-side of Eq.(\ref{e24}) one can obtain $\delta A$ from
Eqs.(\ref{e26'})-(\ref{e26'''}) using in first order $\delta\psi$ by
Eq.(\ref{e10}) and $\delta G$ by Eq.(\ref{e16}).

In Fig.4 we plot the difference $v^p_s(r) - v^p_{0s}(r)$ for the spin
part of the AM operator both in the absence and in the presence of pairing.
As expected, the difference between these two cases is insignificant.
For $\delta V$ the situation is somewhat different. In Fig.5 we plot the
right-hand-side of Eq.(\ref{e24}) for the transitions with $L=0$ (left plot). In
this case the difference is also small. However, for the transitions
with $L=2$ (right plot) the difference is significant. This is the direct
influence of pairing that reduces the transitions to closely lying levels.
The final value of AM does not changed strongly because the contribution
of $X^L[v^p_{0s}]$ with $L=2$ to AM is small.
\begin{figure}
\includegraphics[width=10cm]{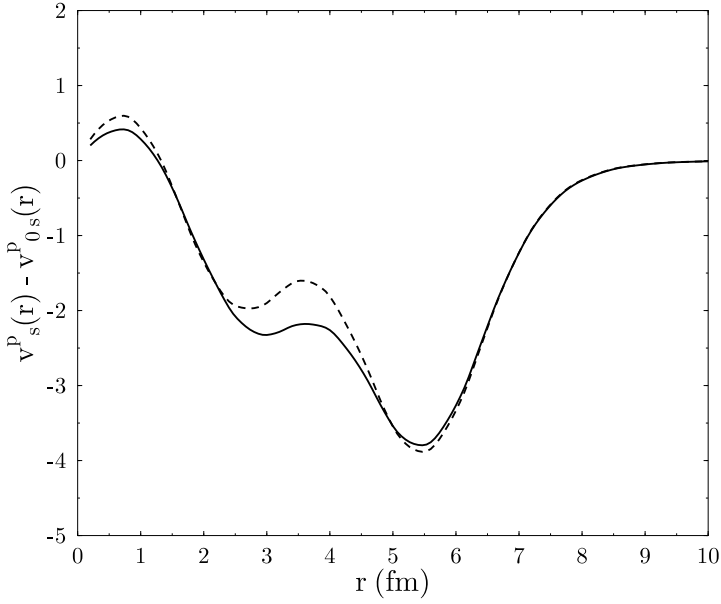}
\caption{The difference $v^p_s(r) - v^p_{0s}(r)$ for the spin part of the AM
operator in the presence of pairing (solid line), and
in the absence of pairing (dashed line).}
\end{figure}
\begin{figure}
\includegraphics[height=5.08cm,width=6cm]{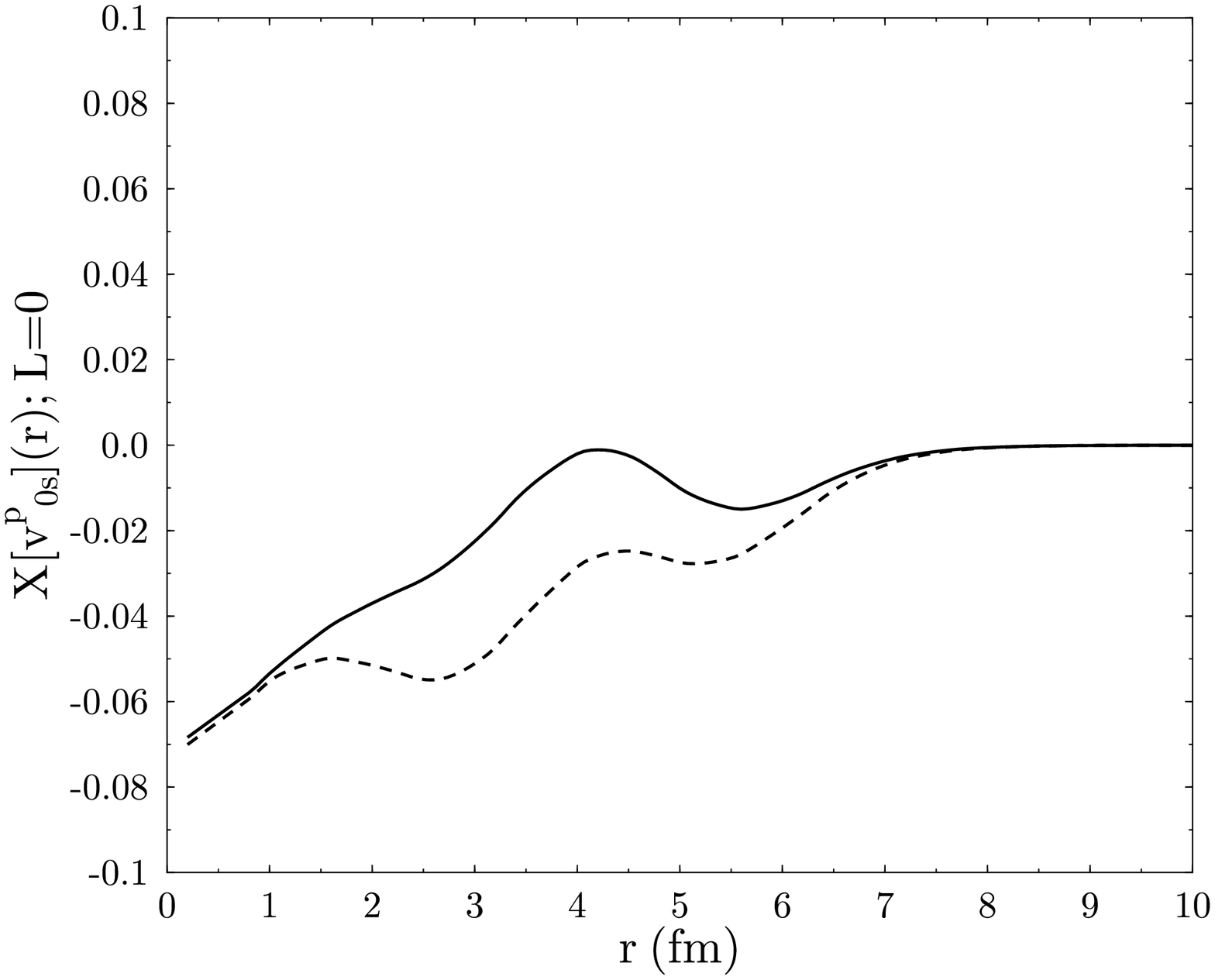}
\includegraphics[height=5cm,width=6cm]{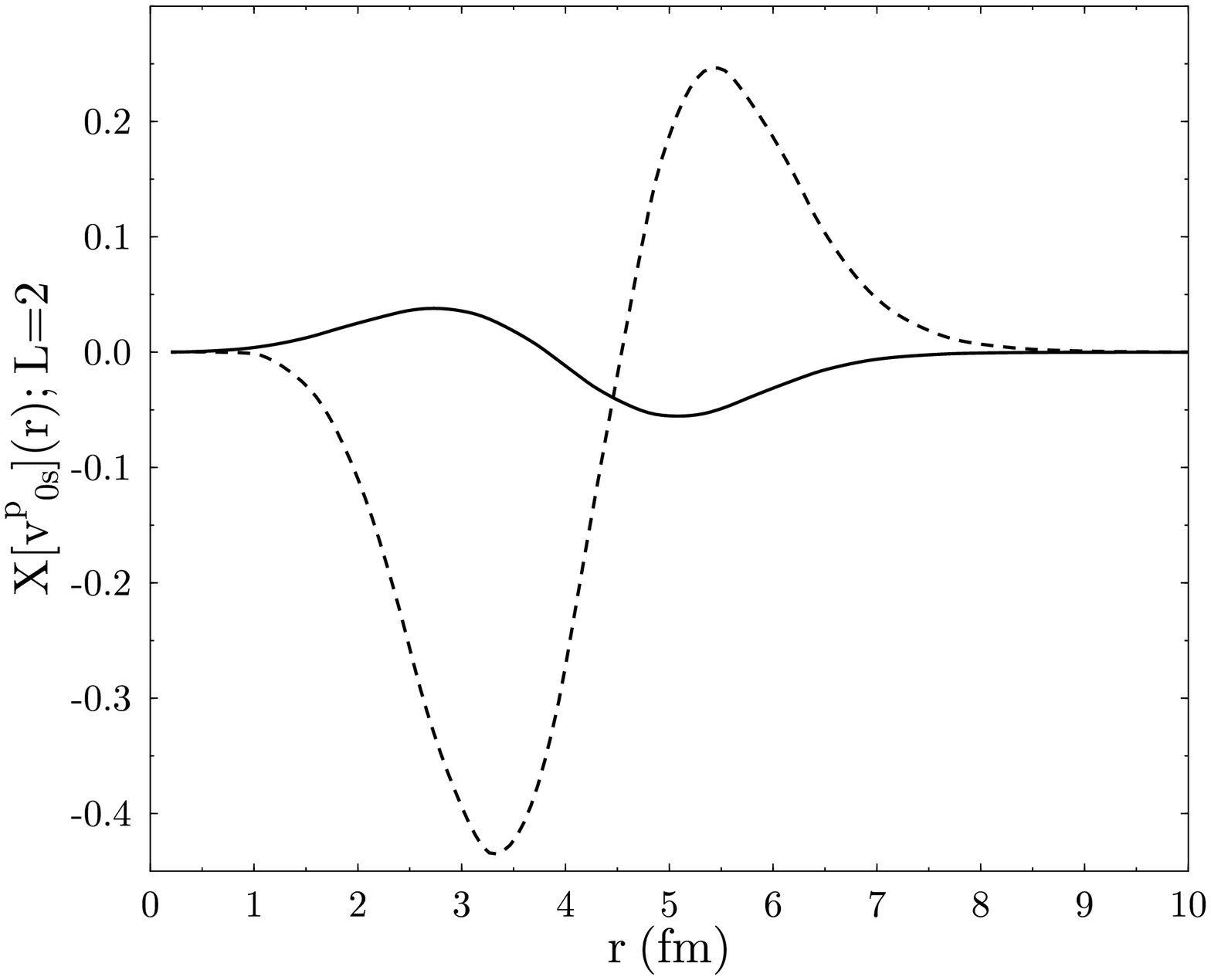}
\caption{The contributions to proton $X^L[v^p_{0s}]$ with different $L$ in
the absence (dashed line), and the presence (solid line) of pairing.
The interaction constant $g_p$ put equal to $1$ for illustrative purpose.
On the left plot $L=0$, on the right plot $L=2$.}
\end{figure}

The results of the anapole moment calculations with pairing are listed in
the Table 1 for $^{133}Cs$ nucleus in the row labeled III.
For the pairing gap $\Delta$ we used the values $\Delta_p=0.95 MeV$
for protons and $\Delta_n=1.05 MeV$ for neutrons from Ref.\cite{MSKH67}.
As expected, the pairing does not
influence much the matrix element of the anapole moment operator $V$. For
$\delta V$ the effects are relatively larger. As mentioned above, the
pairing reduces the transitions near the Fermi surface with small $\Delta E$.
For the $Cs$ nucleus this happens only for the proton transition
$1g_{7/2} - 2d_{5/2}$, $\Delta l=2$ that does not contribute significantly
to the anapole moment.

\section{Parameters of the PNC nuclear forces}
The constants $g_{ab}$ and $g'_{ab}$ of the PNC interaction (\ref{e4}) should
be, strictly speaking, treated as phenomenological ones and should be found
from experimental data. We can, however, try to relate them to the parameters
of the free nucleon-nucleon interaction \cite{DDH}.
This relationship, as obtained in \cite{ST93}, is:
\[
g_{pp} = -(\mu+2)W_\rho A_\rho h^0_\rho
\]
\[
g'_{pp} = g_{nn} = g'_{nn} = g_{pp}
\]
\begin{equation} \label{e27}
g_{pn} = -(2\mu + 1)W_\rho A_\rho h^0_\rho + W_\pi A_\pi f_\pi
\end{equation}
\[
g_{np} = -(2\mu + 1)W_\rho A_\rho h^0_\rho - W_\pi A_\pi f_\pi
\]
\[
g'_{pn} = g'_{np}=(\mu-1)W_\rho A_\rho h^0_\rho,
\]
where
\[
A_\rho =\frac{\sqrt{2}g_\rho}{G_Fm^2_\rho}, \;\;\; A_\pi
=\frac{g_\pi}{G_Fm^2_\pi}
\]
are the dimensional constants, $\mu$ is the isovector nucleon magnetic moment,
and $h^0_\rho$, and $f_\pi$ are the
PNC rho-nucleon and pion-nucleon couplings. The dimensionless factors
$W_\rho$ and $W_\pi$ were introduced to normalize the matrix elements of the
interaction Eq.(\ref{e4}) to the matrix elements of the original finite range
interaction \cite{DDH}.

Following Ref.\cite{WB98}, in Fig. 6 we plotted the extracted values of
coupling constants $H^1_\pi$ and $H^0_\rho$ in the notations of Adelberger
and Haxton \cite{AH85}.
They are related to $f_\pi$ and $h^0_\rho$ by
\[
H^1_\pi= \frac{g_\pi f_\pi}{\sqrt{32}}, \;\;\;\; H^0_\rho = -\frac{g_\rho
h^0_\rho}{2},
\]
\begin{figure}
\includegraphics[width=9cm]{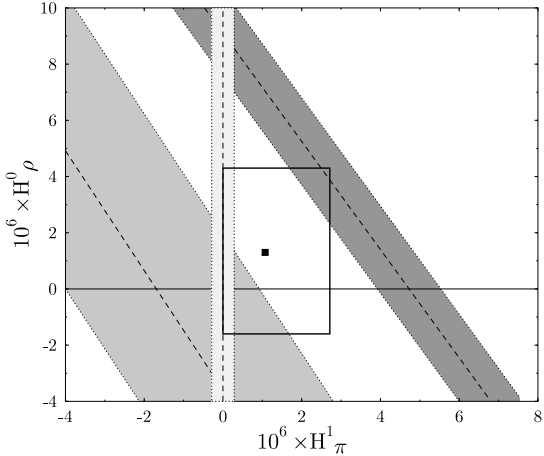}
\caption{Weak coupling constants $H^1_\pi$ and $H^0_\rho$ extracted from
$^{18}F$ (light band), $^{205}Tl$ (medium band), and $^{133}Cs$ (dark band)
experiments. Also shown are the DDH \cite{DDH} "best" values (square) and
"reasonable range" (box)}
\end{figure}
where $g_\pi$ and $g_\rho$ are the strong coupling constants. The bands
corresponding to the $^{205}Tl$ and $^{133}Cs$ data are slightly different
from those extracted in Ref. \cite{WB98}.
There, our previous calculations  \cite{DT97}
were used to extract the coupling constants from the anapole moment data.

Our improved calculations did not change the general situation.
The coupling constants extracted from $^{133}Cs$
and $^{205}Tl$ data still look inconsistent independently on $^{19}F$ data.
This situation eventually raises a question how reliable is the theory of
nuclear anapole moment. Our calculation has been done using the random-phase
approximation. There are two more calculations accounting for many-body
effects \cite{HHM89}, \cite{AB99} within the shell model approach. In Ref.
\cite{AB99} the shell model basis used for calculation of the anapole moment
of $^{205}Tl$ was large enough to account simultaneously both for
the single-particle AM and the core polarization effects. The value obtained
in Ref. \cite{AB99} for the spin part of the anapole moment of $^{205}Tl$ is
$\kappa_s = 0.35$. Our calculation, where we use completely different
quasi-particle interaction and the RPA, gives $\kappa_s = 0.37$ for the "best"
values of the coupling constants. Such close values obtained in completely
different approaches give hope for a weak model dependence of the AM,
although we cannot exclude that this is just the coincidence.

The relation between effective PNC interaction constants and meson-nucleon
coupling constants given by Eq.(\ref{e27}) is less reliable. The
normalization factors $W_\pi$ and $W_\rho$ were determined by comparing
matrix elements of the interaction Eq.(\ref{e4}) and the finite-range DDH
interaction of Ref.\cite{DDH} for $N - {^4He}$ scattering at low energy.
Therefore, we hardly can expect them to be constant in a broad range of all
bound single-particle states. In order to check the accuracy of this
procedure we calculated $W_\pi$
as a ratio of typical matrix elements of the finite range DDH \cite{DDH}
interaction and the zero range interaction Eq.(\ref{e4}). The results are
shown in Fig.7 where the normalization factor $W_\pi$ is plotted as
a function of a single-particle energy for the proton and neutron states.
The factor $W_\pi$ is really state dependent. Only near Fermi surface
$W_\pi = 0.16$. For the states with lower energy it becomes larger reaching
the value $W_\pi \sim 0.4 - 0.5$ for $1s_{1/2}$ states.
\begin{figure}
\includegraphics[width=10cm]{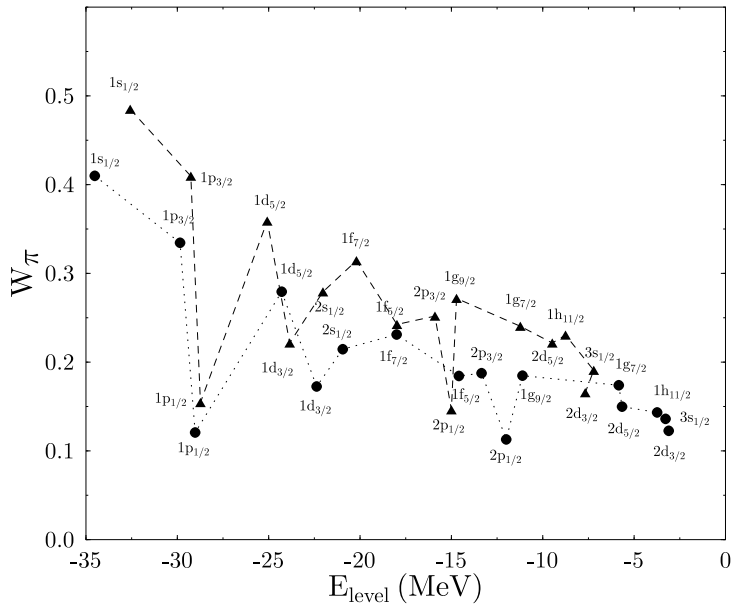}
\caption{Normalization factor $W_\pi$ as a function of bound state energy.
Proton states are connected by dashes. Neutron states are connected by
dots.}
\end{figure}
For this reason one should not use such a simple relation as Eq.(\ref{e27})
for determination of the PNC meson-nucleon coupling constants. There is an
additional reason why Eq.(\ref{e27}) should not be used to extract the
interaction constants. As it was shown in \cite{FV94, FG95} strong
renormalization of the PNC nucleon-nucleon interaction can exist in nuclear
media. As a result, the neutron PNC potential constant $g_n$ (which is small
when estimated using Eq.(\ref{e27})) can be comparable to the proton
constant $g_p$. (See also discussion of the subject in Ref.\cite{D98}.)
From this point of view, the measurements sensitive to $g_n$
would be extremely interesting. These could be the measurements of the anapole
moment of a nucleus with an odd neutron or, another possibility,
the measurement of the neutron spin rotation in helium \cite{DFST83}.

In order to obtain the measured value of the AM of $^{133}Cs$ the PNC
interaction constants Eq.(\ref{e5}) should be increased
approximately by a factor of 2 as compared to their Eq.(\ref{e27}) "best
values". It is interesting to note that a similar conclusion has been
obtained from statistical analysis of the PNC effects
in compound nuclei \cite{TJHB99}.

In Table 3 we show the summarized results for a set of the proton odd and
the neutron odd nuclei.
Here we list just the sum of all contributions. For Ba isotopes the
calculations were performed including the pairing. It is worth noting that for
nuclei with an odd neutron due to the core polarization some contribution
proportional to the proton coupling $g_p$ appears in the anapole moment. In
case of small $g_n$ it can be of the same order as the direct $g_n$
contribution.

\section*{Acknowledgments}
The discussions with I.B. Khriplovich are greatly appreciated. One of the
authors (VD) acknowledges the discussions with B.A. Brown and
V. G. Zelevinsky.
This work was supported by the RFBR grant 98-02-17797.

\renewcommand{\arraystretch}{1.5}

\begin{table}[h]
\caption{Contributions to the anapole moment of $^{133}Cs$}
\begin{center}
\begin{tabular}{||c|c|c|c|c|c||}
\hhline{|t:=:=:=:=:=:=:t|}
&&&&& \\
$\kappa \times 10^2$&s.p.&$V-V_0$&$\delta V$&Total& \\
&&&&& \\
\hhline{|:=:=:=:=:=:=:|}
     &&$-3.6g_p$&$2.5g_p+0.41g_n$&$5.9g_p+0.41g_n$&I \\
\hhline{||~|~|~---||}
$\kappa_s \times 10^2$&$7.0g_p$&&$1.2g_p+0.24g_n$&$4.6g_p+0.24g_n$&II \\
\hhline{||~|~----||}
     &&$-3.5g_p$&$1.9g_p+0.17g_n$&$5.4g_p+0.17g_n$&III \\
\hhline{||------||}
     &&$0.8g_p$&$-0.4g_p-0.03g_n$&$-1.2g_p-0.03g_n$&I \\
\hhline{||~|~|~---||}
$\kappa_{ls} \times 10^2$&$-1.6g_p$&&$-0.3g_p-0.02g_n$&$-1.0g_p-0.02g_n$&II
\\
\hhline{||~|~----||}
     &&$0.8g_p$&$-0.4g_p-0.02g_n$&$-1.2g_p-0.02g_n$&III \\
\hhline{||------||}
     &&$-0.6g_p$&$0.9g_p+0.07g_n$&$-0.2g_p+0.07g_n$&I \\
\hhline{||~|~|~---||}
$\kappa_{conv} \times 10^2$&$-0.5g_p$&&$0.2g_p+0.04g_n$&$-0.9g_p+0.04g_n$&II
\\
\hhline{||~|~----||}
     &&$-0.7g_p$&$-0.1g_p+0.02g_n$&$-1.3g_p+0.02g_n$&III \\
\hhline{||------||}
$\kappa_c \times 10^2$&$0.65g_{pn}$&&&$0.45g_{pn}-0.03g_{np}$&I,II \\
\hhline{||~|~|~|~--||}
     &&&&$0.36g_{pn}-0.02g_{np}$&III \\
\hhline{||------||}
     &&$-3.4g_p$&$2.9g_p+0.45g_n$&$4.4g_p+0.45g_n+\kappa _c$&I \\
\hhline{||~|~|~---||}
$\kappa_{tot} \times 10^2$&$4.9g_p$&&$1.1g_p+0.25g_n$&$2.7g_p+0.25g_n
+\kappa _c$&II \\
\hhline{||~|~----||}
&$+\kappa _c^{s.p.}$&$-3.4g_p$&$1.5g_p+0.18g_n$&$2.9g_p+0.18g_n+\kappa _c$&III
\\
\hhline{|b:=:=:=:=:=:=:b|}
\end{tabular}
\end{center}
\end{table}

\begin{table}[h]
\caption{Contributions to the anapole moment of $^{205}Tl$}
\begin{center}
\begin{tabular}{||c|c|c|c|c|c||}
\hhline{|t:=:=:=:=:=:=:t|}
&&&&& \\
$\kappa \times 10^2$&s.p.&$V-V_0$&$\delta V$&Total& \\
&&&&& \\
\hhline{|:=:=:=:=:=:=:|}
$\kappa _s$&$10.8g_p$&$-5.7g_p$&$4.8g_p+0.14g_n$&$9.9g_p+0.14g_n$&I \\
\hhline{||~|~|~---||}
$\times 10^2$&&&$3.1g_p+0.02g_n$&$8.3g_p+0.02g_n$&II \\
\hhline{||------||}
$\kappa _{ls}$&$-2.2g_p$&$1.1g_p$&$-1.0g_p-0.01g_n$&$-2.0g_p-0.01g_n$&I \\
\hhline{||~|~|~---||}
$\times 10^2$&&&$-0.7g_p-0.01g_n$&$-1.8g_p-0.01g_n$&II \\
\hhline{||------||}
$\kappa _{conv}$&$-0.7g_p$&$-0.9g_p$&$0.9g_p+0.23g_n$&$-0.8g_p+0.23g_n$&I \\
\hhline{||~|~|~---||}
$\times 10^2$&&&$-0.6g_p+0.10g_n$&$-2.2g_p+0.10g_n$&II \\
\hhline{||------||}
$\kappa _c\times 10^2$&$0.85g_{pn}$&&&$0.64g_{pn}-0.06g_{np}$&I,II \\
\hhline{||------||}
$\kappa_{tot}$&$7.8g_p$&$-5.5g_p$&$4.7g_p+0.35g_n$&$7.1g_p+0.35g_n
+\kappa _c$&I \\
\hhline{||~|~|~---||}
$\times 10^2$&$+\kappa _c^{s.p.}$&&$1.9g_p+0.1g_n$&$4.3g_p+0.1g_n
+\kappa _c$&II \\
\hhline{|b:=:=:=:=:=:=:b|}
\end{tabular}
\end{center}
\end{table}

\begin{table}[h]
\caption{Calculated anapole moments for the set of nuclei}
\begin{center}
\begin{tabular}{||c|c|c||}
\hhline{|t:=:=:=:t|}
&& \\
Nuclei &$\kappa _{s.p.}\times 10^2$&$\kappa \times 10^2$ \\
&& \\
\hhline{|:=:=:=:|}
\multicolumn{3}{||c||}{Odd proton nuclei} \\
\hhline{||---||}
$^{133}$Cs&$4.9g_p+0.65g_{pn}$&$2.9g_p+0.18g_n+0.36g_{pn}-0.02g_{np}$ \\
\hhline{||---||}
$^{205}$Tl&$7.8g_p+0.85g_{pn}$&$4.3g_p+0.1g_n+0.64g_{pn}-0.06g_{np}$ \\
\hhline{||---||}
$^{209}$Bi&$5.4g_p + 0.96g_{pn}$&$ 2.5g_p + 0.3g_n + 0.57g_{pn} - 0.04g_{np}$
\\
\hhline{||---||}
\multicolumn{3}{||c||}{Odd neutron nuclei} \\
\hhline{||---||}
$^{135}$Ba&$-6.5g_n - 0.25g_{np}$&$-0.1g_p - 4.6g_n + 0.01g_{pn} - 0.19g_{np}$
\\
\hhline{||---||}
$^{137}$Ba&$-6.5g_n - 0.25g_{np}$&$-0.2g_p - 5.7g_n + 0.01g_{pn} - 0.23g_{np}$
\\
\hhline{||---||}
$^{207}$Pb&$-9.6g_n - 0.16g_{np}$&$-0.1g_p - 6.7g_n + 0.01g_{pn} - 0.14g_{np}$
\\
%\hhline{||---||}
%$^{209}$Pb&$-5.2g_n - 0.44g_{np}$&$-0.2g_p - 4.6g_n + 0.03g_{pn} - 0.29g_{np}$
\\
\hhline{|b:=:=:=:b|}
\end{tabular}
\end{center}
\end{table}

\end{document}